\documentclass{ckm}                 

\confname{Workshop on the CKM Unitarity Triangle, IPPP Durham, April
  2003}

\title{B Physics triggers at CMS}

\author{A. Starodumov\thanks{On leave from ITEP, Moscow, Russia}, on behalf
of the CMS Collaboration}

\address{INFN, Sezione di Pisa, Italy}

\begin{document}

\begin{abstract}
The CMS detector is mainly 
designed to investigate hard events.
Only few Level-1 Trigger conditions are suitable to select soft B-meson decays.
The B-physics potential of CMS depends strongly on a selection
strategy at 
High-Level Trigger. The selection algorithms for some benchmark B-decay
channels that allow CMS to perform competitive B-physics 
program are presented. 
\end{abstract}

\maketitle


\section{Introduction}


The CMS detector~\cite{cms} is 
designed and optimized for discovery physics, 
in particular the observation of the Higgs boson(s) and the investigation 
of supersymmetry or other physics beyond the standard model. 
Nevertheless, 
B physics
may be an interesting subject, especially during the low luminosity 
(2$\times$10$^{33}$ cm$^{-2}$ s$^{-1}$) phase.
The study presented 
in this note has been done for this phase.

The B-physics potential of the CMS detector
 depends on many factors but mostly 
on the trigger conditions that will 
be used to collect events for further offline analysis. 
While the main Level-1 triggers are almost defined,
the High-Level Triggers (HLT) are currently 
under discussion and investigation.
In this note, possible HLT selection algorithms for several B decay channels 
of interest are presented. 

\section{The CMS Trigger and Data Acquisition}

The CMS  Trigger and Data Acquisition System (TriDAS) is designed to inspect
the detector information at the full beam crossing rate of 40 MHz and
to select events at a maximum rate of O($10^2$) Hz for archiving and later 
offline analysis. The required rejection power of O($10^5$) is too large 
to be archived in a single processing step. For this reason, the full 
selection task is split in two steps. The first step (Level-1 Trigger)
is designed to reduce the rate of events accepted for further processing 
to less than 100 kHz. The time available for actually processing the data 
is about 1$\mu$s. Therefore the Level-1 Trigger can only process data
from a subset of CMS sub-detectors, the calorimeters and muon chambers. 
Moreover, this data do not represent the full information recorded in the
CMS front-end electronics, but only a coarse-granularity and lower-resolution 
set.
The second step (High-Level Trigger) is designed to
reduce the maximum Level-1 accept rate of 100 kHz to a final 
output rate of 100 Hz. The event selection at 
this stage takes advantage of the enhanced granularity and resolution 
available and makes use of information from tracking detectors. The HLT 
selection algorithms are actually 
almost as sophisticated as those used in the offline
reconstruction of the events. The current best estimate for the total
average processing time required for the design selection of 1:1000 
(i.e. a rate reduction from 100 kHz to 100 Hz) is roughly
40 ms, with some events requiring up to one second.

The Level-1 Trigger is documented in the Volume I of the
TriDAS Technical Design Report~\cite{vol1} and details of the 
architecture of the TriDAs related to the HLT can be found 
in the Volume II~\cite{vol2}.

It is important to mention that the current CMS plan 
foresees only partial completion of the Data Acquisition
 system at the low luminosity phase
(mainly during which the B-physics study will be performed).  
At startup, the DAQ system will therefore only be able to
 handle an event rate of 50 kHz or less. Moreover,
to account for all uncertainties in the simulation of the basic physics
processes, the CMS detector and the beam conditions, an additional 
safety factor of 3 is taken into account. Therefore, the maximum
Level-1 Trigger rate of 16 kHz is considered for the HLT study.
  
In the following, the CPU time required for the algorithm execution at the 
HLT is also shown. This time is normalized on a Pentium-III 1 GHz
processor and fits within the foreseen time budget if it is of the order of 
500 ms/event ~\cite{vol2}.

\section{The Level-1 Triggers}

The allocation of the Level-1 Trigger bandwidth has been optimized in order to 
ensure the widest possible physics reach for the CMS experiment. An equal
allocation (4 kHz) across the four categories of "objects", taken as 
1) electrons/photons, 2) muons, 3) tau jets (hadronic tau decays) and 
4) jets plus combined channels (at least one jet plus other Level-1
signatures with lower thresholds),
is made.  The thresholds for each trigger were 
set to satisfy this requirement. 
Since the CMS detector is mainly oriented towards 
the investigation of events
with high $p_T$ muons and electrons, very energetic jets and photons or 
significant missing $E_T$ values,
the only suitable triggers which can collect reasonable amount of
b events are the single-muon ($p_T>$14 GeV/$c$), di-muon ($p_T>$3 GeV/$c$) 
and possibly single-muon-plus-jet triggers. 

\section{Benchmark channels}

According to the Level-1 triggers
the B-meson-decay final states should contain one or two
relatively high $p_T$ muon(s). Based on this requirements
the following three benchmark channels have been selected:

\begin{equation}
 {\rm B^0_s} \rightarrow  {\rm J}/\Psi \phi \rightarrow \mu^+ \mu^-  
{\rm K^+K^-};
\label{eq1}
\end{equation}
\begin{equation}  {\rm B^0_s}\rightarrow \mu^+ \mu^-;
\label{eq2}
\end{equation}
\begin{equation}  {\rm B^0_s}\rightarrow  {\rm D_s^-} \pi^+ \rightarrow
\phi \pi^- \pi^+ \rightarrow  {\rm K^+K^-} \pi^- \pi^+.
\label{eq3}
\end{equation}

The channels (\ref{eq1}) and (\ref{eq2}) are triggered by the 
di-muon Level-1 trigger and the channel (\ref{eq3}) by the 
single-muon (which comes from the decay of the second B-meson 
in the event) Level-1 trigger. 

\section{High-Level Trigger strategy}

\begin{figure}[htbp]
\hbox to\hsize{\hss
\includegraphics[width=\hsize]{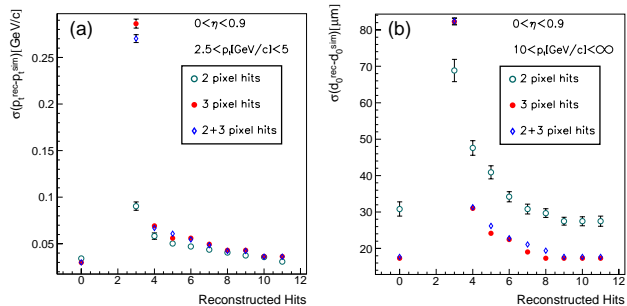}
\hss}
\caption{The resolution on a) $p_T$ and b) impact parameter for partial track
reconstruction, compared with full track resolution, as a function of the 
number of smoothing steps in different $p_T$ and for the barrel region. 
The leftmost points at "0" reconstructed hits show the full tracker 
performance.}
\label{fig:mom-res}
\end{figure}

The HLT strategy is common for all three 
benchmark channels. The CMS Tracker~\cite{tracker} information is used 
to reconstruct tracks,
compute invariant masses and select events on the basis of kinematical and 
topological properties of the final states of the benchmark channels. 
A difference between
algorithms is related to the track reconstruction procedure. Due to
 the physically limited HLT processing time,  
it is not possible to reconstruct all
tracks with $p_T>$1 GeV/$c$
 in the full rapidity coverage.
Nevertheless, for the channels (\ref{eq1}) and (\ref{eq2}) there is a natural, 
 so-called, Region of Interest (RoI) around the
two Level-1 muons. Moreover, the large magnetic field 
  yields a satisfactory momentum 
resolution for these 
muons even with a limited number of hits.  The resulting accuracy 
of the track parameters obtained with a partial track 
reconstruction is shown in 
 Fig. \ref{fig:mom-res}. The three cases shown in this figure differ by the 
number of layers hit in the pixel detector used in the reconstruction:
1) only two innermost layers hit; 
2) all three layers hit; 3) at least two
layers hit.  
For the channels (\ref{eq1}) and (\ref{eq2}), 
the  track reconstruction is
performed in the RoI and stopped if one of the conditions is fulfilled: 
1) either six hits along the trajectory are found, or
2) the relative transverse momentum resolution is better then 2\%.
For the channel (\ref{eq3}), the direction of the Level-1-Trigger muon is not
correlated with the direction of the $ {\rm B^0_s}$~\cite{bbprod}.
Hence, the  reconstruction of the $ {\rm B^0_s}$-decay products has to be done
 in the full tracker volume. To stay in the HLT time budget, the 
track reconstruction was performed with only 
three hits (two pixel and one innermost strip) along the trajectory. 
Because the majority of tracks 
in this decay mode
have  a transverse momentum not larger than 2 GeV/$c$, a momentum resolution
of 2\% can be achieved~\cite{hlttrack} with only three hits.  
  

\subsection{The ${\rm B^0_s} \rightarrow  \mu^+  \mu^-$ channel}

Events are required to have two oppositely charged isolated muons coming 
from a common displaced vertex with an invariant mass close to the 
nominal ${\rm B^0_s}$ mass. First, a fast algorithm
creates track seeds from pairs of hits in the pixel detector
 which can be assigned to tracks with $p_T >$4 Gev/$c$ with transverse
impact parameter smaller than 1 mm. The reconstructed track pairs are used
for a first reconstruction of the primary vertex. 
Then, pairs of hits are filtered using a vertex 
constraint and the RoI around the Level-1 muon momentum directions 
($\Delta \eta \leq$ 0.5
 and $\Delta \phi \leq$ 0.8). 
These pairs are used for the conditional track reconstruction
as described above.

The reconstructed tracks are propagated back to the origin using a standard 
smoothing procedure~\cite{vol2}. 
If and only if two oppositely charged particle tracks are 
reconstructed, it is checked that the invariant mass formed by 
the two particles is within 
150 MeV/$c^2$ of the ${\rm B^0_s}$ mass. 
In order to suppress combinatorial
background, tracks are vertexed and retained if the vertex 
$\chi^2$ is smaller than 20
and the decay length in the transverse plane is larger than 150 $\mu$m. 
The HLT 
selection efficiency is about 33\% and the global efficiency 
(which included the Level-1-Trigger efficiency) 
is about 5\%. The time needed on average for such a 
reconstruction is about 240 ms.
About 94 signal events are expected to be selected by this algorithm in
a low-luminosity year.
The background rate after HLT is below 2 Hz.

\subsection{The ${\rm B^0_s} \rightarrow {\rm J}/\Psi \phi$ channel}

Muons from J/$\Psi$ are reconstructed in the same way as described 
above,
except for 
the fact that slightly tighter cuts on the 
di-muon mass and the secondary vertex are put to keep the background
rate at an acceptable level.
The mass window is decreased to 100 MeV/$c^2$ and the secondary vertex has to 
be at least 200 $\mu$m from the beam with the vertex $\chi^2 <$10.
The di-muon selection leads to an inclusive rate of about 15 Hz, 
90\% is due to a J/$\Psi$ from b decays. 
 
\begin{figure}[htbp]
\hbox to\hsize{
\hss
\includegraphics[width=0.55\hsize]{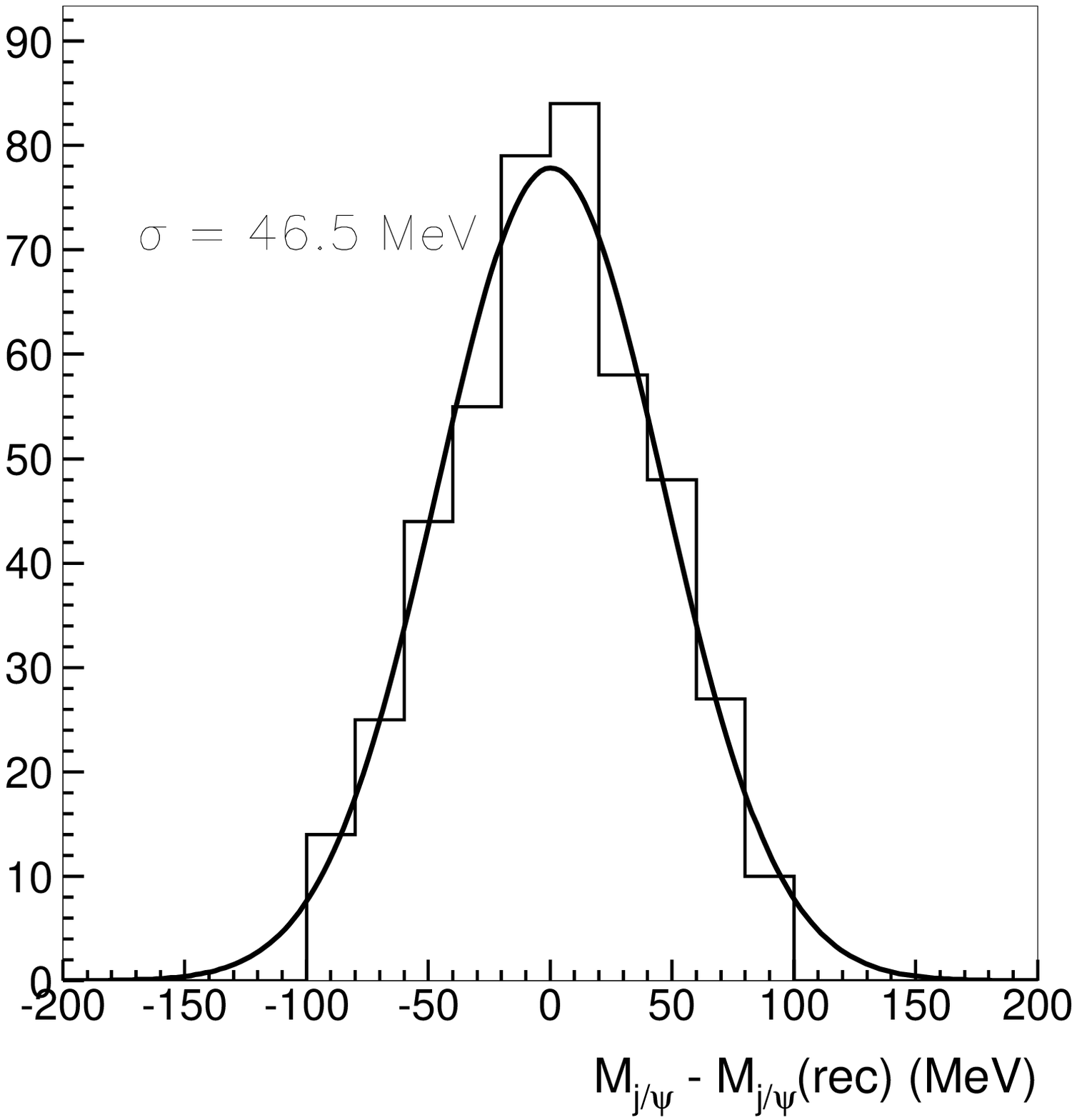}
\hss
\includegraphics[width=0.55\hsize]{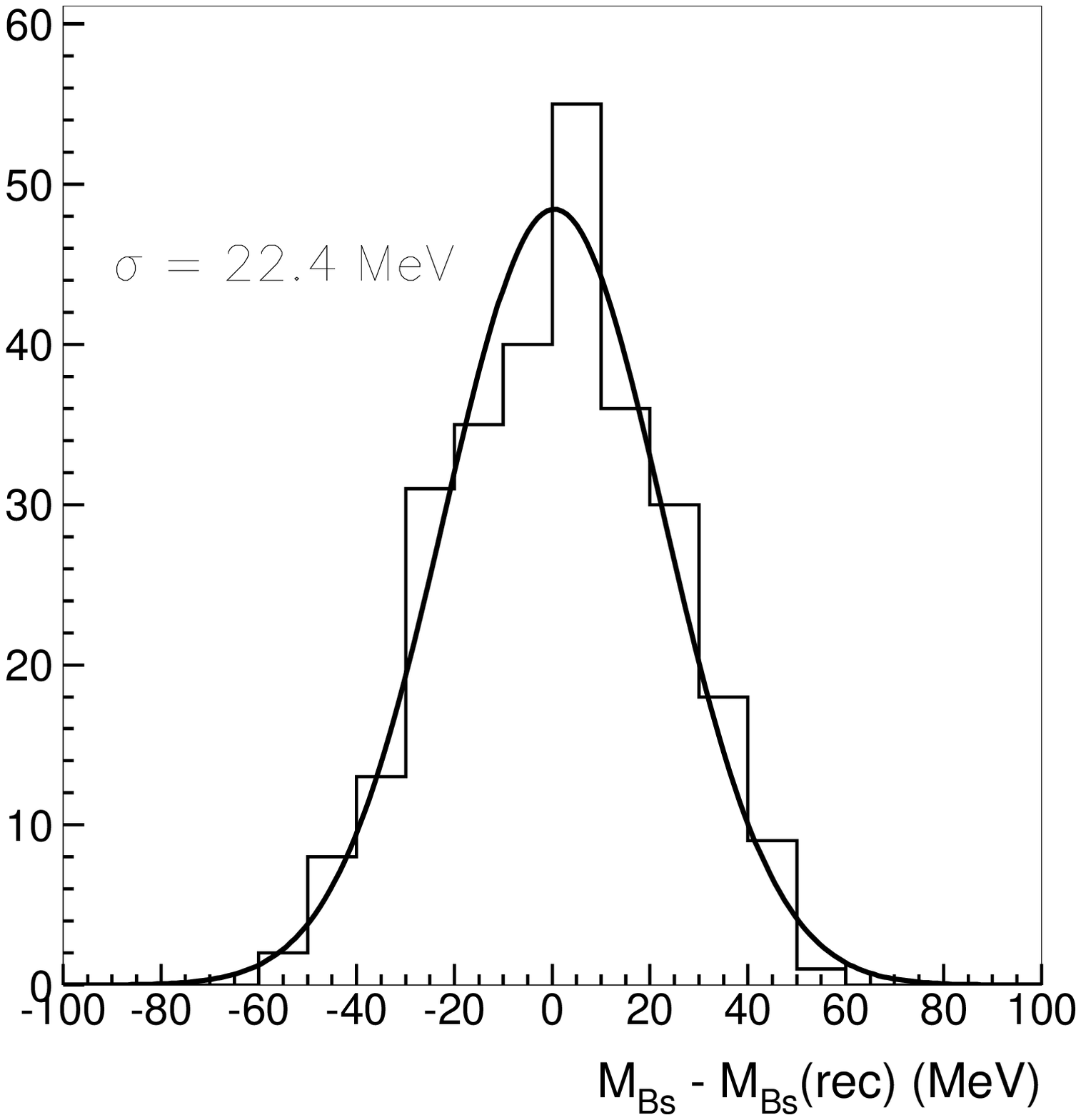}
\hss}
\caption{Mass resolutions for J/$\Psi$ and ${\rm B^0_s}$ candidates for
the ${\rm B^0_s} \rightarrow {\rm J}/ \Psi \phi$ decay channel.}
\label{fig:mbs}
\end{figure}

To select the kaons, coming from the $\phi$ meson, all track 
seeds within a cone $\Delta R <$1.5 around the J/$\Psi$ candidate
direction are reconstructed using the conditional track reconstruction. 
Charged-particle tracks of opposite sign are 
paired and retained if their invariant mass is within 10 MeV/$c^2$ of the 
$\phi$ mass. 
The ${\rm B^0_s}$ 
candidate is retained if the mass formed by the  J/$\Psi$ and $\phi$ 
candidates is
 within 60 MeV/$c^2$ of the ${\rm B^0_s}$ mass. 
The mass resolutions for the J/$\Psi$ and the
${\rm B^0_s}$ candidates are shown in Fig.~\ref{fig:mbs}.

The global HLT selection efficiency  
is about 8.7\% and the yield is  about 160000 events for
a low-luminosity year.
The background rate is
estimated to be less than 2 Hz. 
The average execution time for signal and background events 
is about 260 ms for the reconstruction 
of the J/$\Psi$ and 800 ms for the full ${\rm B^0_s}$ reconstruction. 

\subsection{The ${\rm B^0_s} \rightarrow {\rm D_s^-} \pi^+$ channel}

A single 
muon or a combination of a low $p_T$ muon and a relatively low $E_T$ jet 
can be used as a Level-1 trigger to select this decay channel. 
Table~\ref{tab:bkg} shows the expected trigger rates as a function of the 
different cuts on the muon and the jet transverse momentum and energy. 
The corresponding yields
for signal events are shown in Table~\ref{tab:sig}.  
The machine conditions and the instantaneous luminosity may well allow 
the Level-1 Trigger to be run with lower thresholds than the nominal ones, 
which have been determined for the luminosity of 
2$\times$10$^{33}$ cm$^{-2}$ s$^{-1}$. 
\begin{table}[htbp]
 \begin{center}
 \begin{tabular}{||c||c|c||}  \hline
{\small $p_T^{\mu}$} & \multicolumn{2}{c||}{\bf Background rate (Hz)}\\ 
\cline{2-3}
{\small (GeV/$c$)}&{\small No jet}& 
 {\small $E_T^{jet} \ge$30 GeV}\\ \hline \hline 
 4& 270 (50K)& 80 (5.7K)\\ \hline
 7 & 110 (18K)&  50 (2.7K)\\ \hline
10 & 40 (6.4K)&  10 (1.3K)\\ \hline
14 & 17 (3.2K)&   8 (0.7K)\\ \hline
\end{tabular}
\caption{\small HLT (Level-1) rates as a function
of trigger $\mu$ momentum and trigger jet $E_T$ for an integrated luminosity
of 20 fb$^{-1}$.}
\label{tab:bkg}
\end{center}
\end{table}

\begin{figure}[htbp]
\hbox to\hsize{\hss
\includegraphics[width=0.55\hsize]{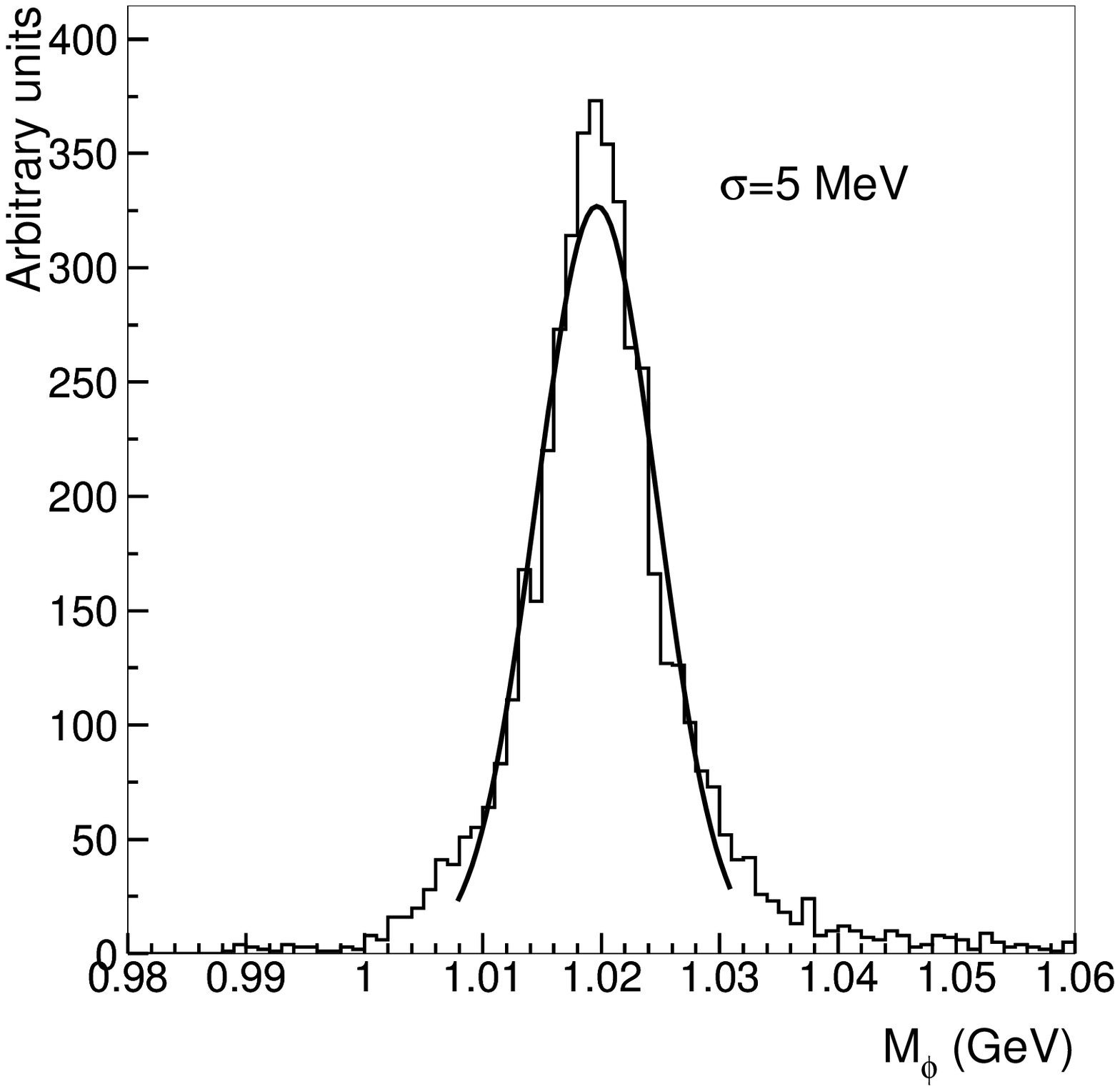}
\hss
\includegraphics[width=0.55\hsize]{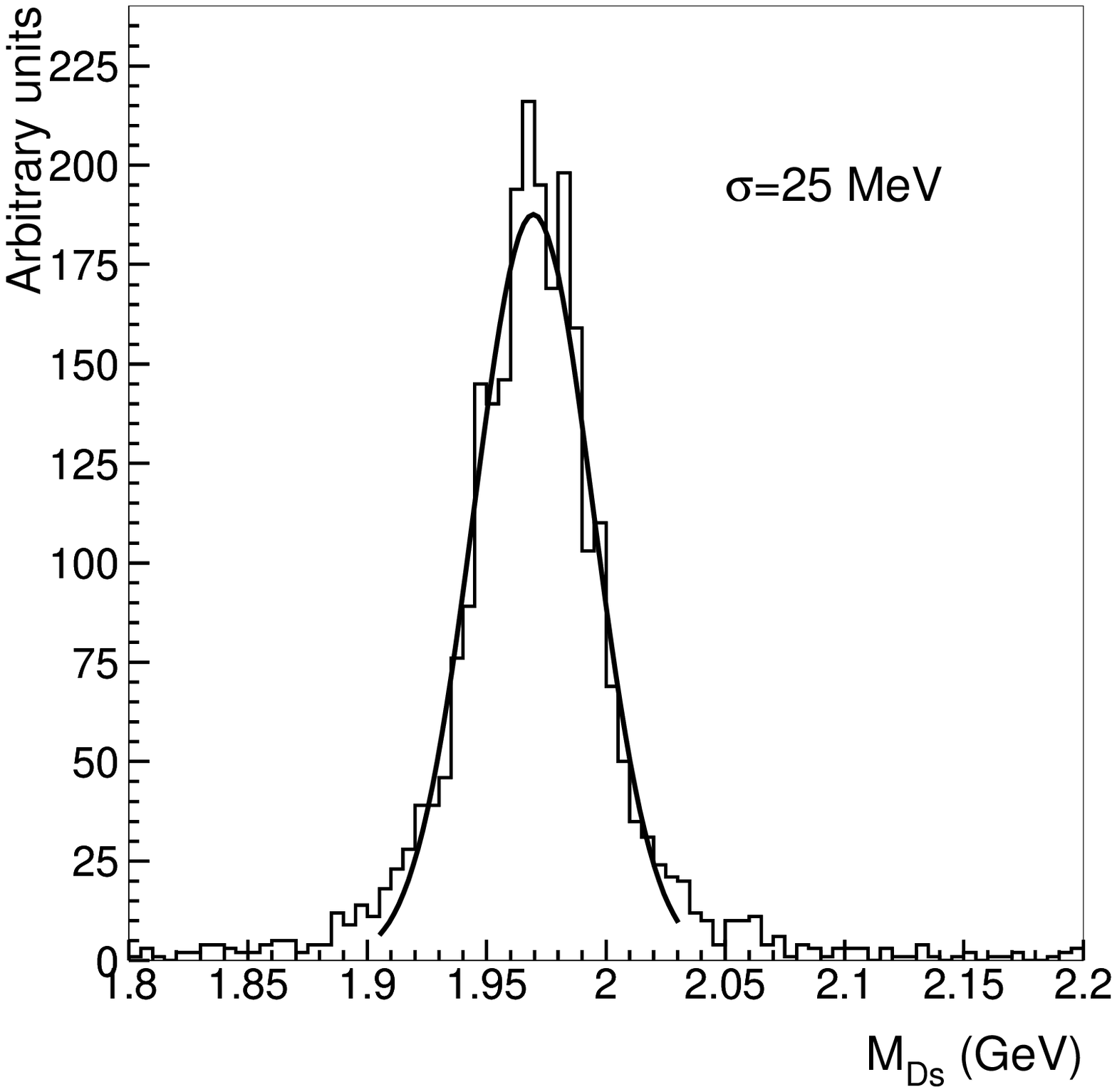}
\hss}
\caption{Invariant mass distribution for $\phi$ and ${\rm D_s}$ candidates from
${\rm B^0_s} \rightarrow {\rm D^-_s} \pi^+ \rightarrow \phi \pi^- \pi^+$ 
decay channel.}
\label{fig:mbsds}
\end{figure}

\begin{figure}[htbp]
\hbox to\hsize{\hss
\includegraphics[width=0.55\hsize]{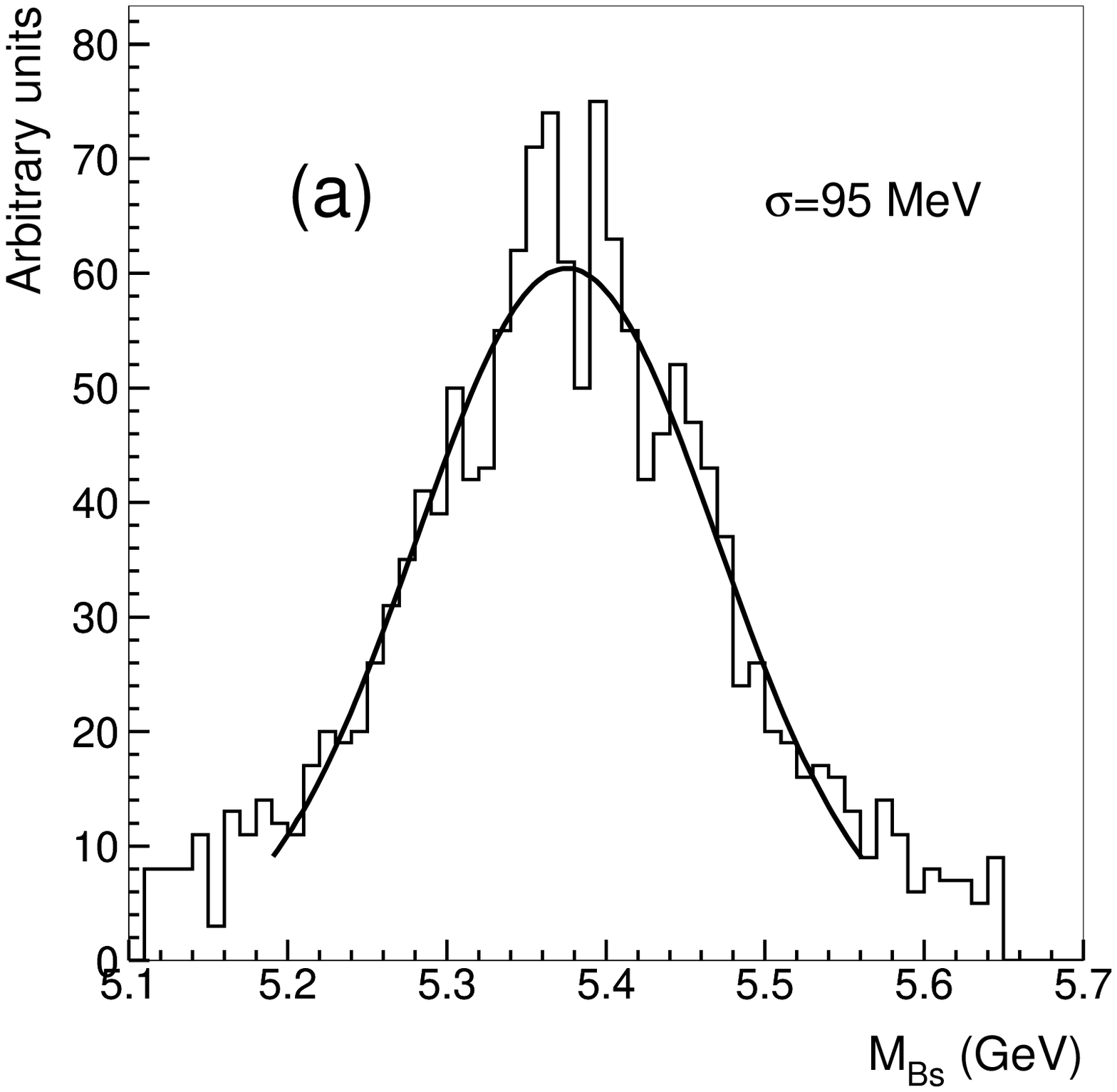}
\hss
\includegraphics[width=0.55\hsize]{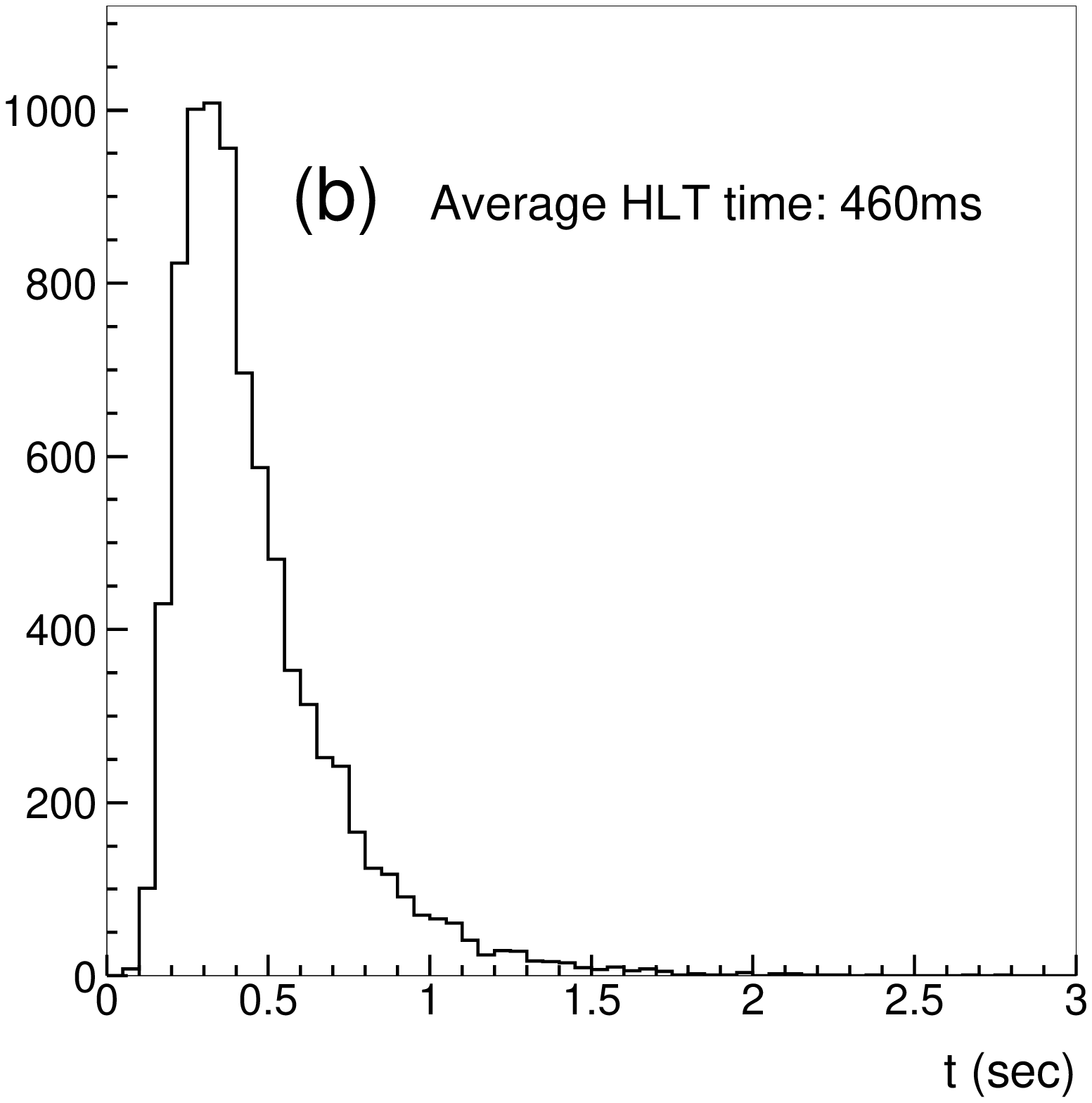}
\hss}
\caption{Invariant mass of ${\rm B^0_s}$ candidate (a) and HLT selection 
time 
(b) for the decay channel ${\rm B^0_s} \rightarrow {\rm D^-_s} \pi^+$.}
\label{fig:bstime}
\end{figure}

First, the primary vertex is reconstructed based on the pixel information. 
Then, all tracks with $p_T >$0.7 GeV/$c$ are 
reconstructed from three hits only:
two hits are taken from the pixel detector and one from the first silicon 
strip layer.
\begin{table}[htbp]
 \begin{center}
 \begin{tabular}{||c||c|c||}  \hline
{\small $p_T^{\mu}$} & \multicolumn{2}{c||}{\bf Number of Signal events} \\ 
\cline{2-3} {\small (GeV/$c$)}&{\small No jet}& {\small $E_T^{jet} \ge$30 GeV} \\ \hline \hline 
 4& 11K (106K)& 2.3K (14K)\\ \hline
 7 &  5.6K (48K)& 1.3K (9.4K)\\ \hline
10 &  2.0K (20K)& 0.6K (5.3K)\\ \hline
14 &  1.0K (11K)&  0.3K (3K)\\ \hline
\end{tabular}
\caption{\small Number of signal events after HLT (Level-1) as a function
of trigger $\mu$ momentum and trigger jet $E_T$ for an integrated luminosity
of 20 fb$^{-1}$.}
\label{tab:sig}
\end{center}
\end{table}
To reduce the
execution time, seeds are requested to come from the primary vertex $z$ 
location within $\pm$1 mm. 
A search is then made for the $\phi$, ${\rm D_s}$ and ${\rm B^0_s}$ 
candidates using following topological requirements: 
$\Delta R ({\rm K,K}) <$0.3, 
$\Delta R (\phi, \pi) <$1.2 and 
$\Delta R ({\rm D_s}, \pi) <$3.0. 
The invariant mass distributions are shown
in  Fig.~\ref{fig:mbsds}
for the $\phi$ and the ${\rm D_s}$ 
candidates and in Fig.~\ref{fig:bstime}a for the ${\rm B^0_s}$ candidate. 
The main cuts are placed on the invariant masses 
and the transverse momenta of the
three resonances. A three-dimensional reconstruction of 
the ${\rm D_s}$ decay vertex is also performed.

The signal efficiency after all cuts is about 10\%, while the background 
is suppressed by a factor of 250. The mean execution time
 shown in Fig.~\ref{fig:bstime}b is about 460 ms for signal events 
and about 640 ms for background. 
The number of signal events depends on the  
bandwidth allocated to this channel, according to the instantaneous luminosity 
of LHC. From the above tables even a relatively high threshold of 14 GeV/$c$ 
gives a rate of 3 kHz after Level 1. Since the 
allowed bandwidth may be lower 
than 3 kHz, it is necessary to scale the rate. This can be done 
by scaling either 
the rate of events with the lowest possible $p_T$ threshold of the trigger
muon or the rate of events in all  
$p_T$ bins so that the overall sum matches the allowed bandwidth. 
On the other hand, the quality of the analysis depends on the total 
number of events, the fraction of signal events and their dilution factor. 
By rising the muon $p_T$ the signal fraction and the dilution factor increases 
mildly, while the total number of events decreases. Therefore, it is better 
to fill the allowed bandwidth with events with the lowest possible muon 
$p_T$ and, if it is needed, downscale only the next lower bin.
For example, referring to Table~\ref{tab:bkg}, 
for an allowed bandwidth of 4 kHz, 
all events with a muon of $p_T >$14 GeV/$c$ should be kept, while events with 
10 GeV/$c < p_T <$14 GeV/$c$  should be scaled by a factor of 4. 
Details of this study can be found in~\cite{bsds}.
As mentioned above, the number of signal events collected per one year 
depends on the bandwidth allocated for this decay channel
at Level 1. 
For a Level-1-Trigger rate of 1 KHz the number of selected signal events
ranges from 300 to 400 for the integrated luminosity of 20 fb$^{-1}$ 
and depends on the type of the Level-1 trigger. 
With such an amount of events CMS will be sensitive up to 
$\Delta m_s$=20 ps$^{-1}$ \cite{xie}.
\section{Conclusion}

Although B physics at a general-purpose detector like CMS is not a 
first priority task, a proper design of the sub-detectors and the TriDAS system
allow CMS to be competitive for a number of  
B-decay channels. Single-muon 
and di-muons final states of B decays can be successfully triggered
both at Level-1 and HLT, with enough signal events for the offline
analysis. 
The HLT algorithms presented in this note are based on the CMS Tracker
information used to perform a fast and precise primary/secondary 
vertex, track and invariant mass reconstruction. 
All algorithms are channel specific and very similar to the offline ones 
with slightly looser cuts.
The use of the 
conditional track reconstruction in a Region of Interest or the track 
reconstruction in the full detector acceptance with only three innermost 
hits allows the algorithms to be fast enough to fit within the HLT time
budget.

More studies will be done to optimize the selection procedure for 
the decay mode ${\rm B^0_s} \rightarrow {\rm D^-_s} \pi^+$ and to decrease 
the HLT execution
time for the decay mode ${\rm B^0_s} \rightarrow {\rm J}/ \Psi \phi$.

\end{document}